\documentclass[journal=jctcce,manuscript=article,etalmode=truncate,maxauthors=0]{achemso}
\setkeys{acs}{articletitle = true}

\usepackage{graphicx}
\usepackage{dcolumn}
\usepackage{bm}

\usepackage[version=3]{mhchem}
\usepackage[utf8]{inputenc}
\usepackage[T1]{fontenc}
\usepackage{mathptmx}
\usepackage{booktabs}
\usepackage{amsmath}
\usepackage{caption}
\usepackage{subcaption}
\usepackage{makecell}
\usepackage[flushleft]{threeparttable}

\usepackage{xcolor}
\usepackage{lmodern} 
\usepackage{dcolumn}
\SectionNumbersOn

\author{Till Kirsch}
\affiliation{Department Chemie, Johannes Gutenberg-Universit\"at Mainz, Duesbergweg 10-14, D-55128 Mainz, Germany}
\email{kirsch@uni-mainz.de}

\author{J\'{o}gvan Magnus Haugaard Olsen}
\affiliation{DTU Chemistry, Technical University of Denmark, DK-2800 Kgs. Lyngby, Denmark}

\author{Viacheslav Bolnykh}
\affiliation{Laboratory of Computational Chemistry and Biochemistry, École Polytechnique Fédérale de Lausanne, CH-1015 Lausanne, Switzerland}

\author{Simone Meloni}
\affiliation{Dipartimento di Scienze Chimiche, Farmaceutiche ed Agrarie, Universita degli Studi di Ferrara,  44121 Ferrara, Italy}

\author{Emiliano Ippoliti}
\affiliation{Institute for Advanced Simulation (IAS-5) and Institute of Neuroscience and Medicine (INM-9), Forschungszentrum J\"ulich, 52425
J\"ulich, Germany}

\author{Ursula Rothlisberger}
\affiliation{Laboratory of Computational Chemistry and Biochemistry, École Polytechnique Fédérale de Lausanne, CH-1015 Lausanne, Switzerland}

\author{Michele Cascella}
\affiliation{Department of Chemistry and Hylleraas Centre for Quantum Molecular Sciences, University of Oslo, PO Box 1033 Blindern,  0315 Oslo, Norway}

\author{J\"urgen Gauss}
\affiliation{Department Chemie, Johannes Gutenberg-Universit\"at Mainz, Duesbergweg 10-14, D-55128 Mainz, Germany}

\title{Wavefunction-based electrostatic-embedding QM/MM using CFOUR through MiMiC}

\newcommand{\be}{\begin{equation}}
\newcommand{\ee}{\end{equation}}
\newcommand{\hH}{\hat{H}}

\newcommand{\bs}{\boldsymbol}
\newcommand{\bR}{\boldsymbol{R}}
\newcommand{\br}{\boldsymbol{r}}

\begin{document}  

\newpage

\begin{abstract}
We present an interface of the wavefunction-based quantum-chemical software CFOUR to the multiscale modeling framework MiMiC. Electrostatic embedding of the quantum-mechanical (QM) part is achieved by analytic evaluation of one-electron integrals in CFOUR, while the rest of the QM/MM operations are treated according to the previous MiMiC-based QM/MM implementation. Long-range electrostatic interactions are treated by a multipole expansion of the potential from the QM electron density to reduce the computational cost without loss of accuracy. Testing on model water/water systems, we verified that the CFOUR interface to MiMiC is robust, guaranteeing fast convergence of the SCF cycles and optimal conservation of the energy during the integration of the equations of motion. Finally, we verified that the CFOUR interface to MiMiC is compatible with the use of a QM/QM multiple time-step algorithm, which effectively reduces the cost of AIMD or QM/MM-MD simulations using higher level wavefunction-based approaches compared to cheaper density-functional theory-based ones. The new wavefunction-based AIMD and QM/MM-MD implementation was tested and validated for a large number of wavefunction approaches, including Hartree-Fock and post-Hartree-Fock methods like M\o{}ller-Plesset, coupled cluster, and complete active space self-consistent field.
\end{abstract}

\clearpage

\section{Introduction}

Multiscale modeling techniques that involve quantum-chemical methods\cite{szabo} are important tools widely used in many areas such as solution chemistry, catalysis, or enzymology.\cite{warshel, senn, brunk, tomasi1,tomasi2} Among them, hybrid quantum mechanics/molecular mechanics (QM/MM) approaches are especially important in reactive biochemical systems where the explicit treatment of the electronic structure as well as the handling of the environment, including several tens to hundreds of thousands of atoms, is mandatory.\cite{senn2, friesner, campomanes, campomanes2} Regardless of the progress in linear scaling\cite{zalesny} and parallelization of electronic-structure methods,\cite{skylaris} a fully quantum-mechanical (QM) treatment for such intrinsically large systems is not possible. Fortunately, a proper description of the electronic structure by quantum-chemical methods is only required for smaller regions of the system where, e.g., chemical processes take place. In turn, all the other parts of the system can be described effectively by some simplified approach, typically at the molecular-mechanical (MM) level by a classical force-field.\cite{ponder, oostenbrink, jorgensen} 

In additive QM/MM approaches\cite{brunk} the Hamiltonian of the whole system is split into three parts:
\be
\hH_{\mathrm{tot}} = \hH_{\mathrm{QM}} + \hH_{\mathrm{MM}} + \hH_{\mathrm{QM/MM}}.
\ee
While $\hH_{\mathrm{QM}}$ and $\hH_{\mathrm{MM}}$ are the Hamiltonians of the QM and MM subsystem, the crucial part of any QM/MM implementation is a proper description of the coupling Hamiltonian $\hH_{\mathrm{QM/MM}}$ of the two subsystems. The QM/MM coupling is usually done at one of three levels of complexity.\cite{senn, brunk} The simplest one is {\em mechanical embedding},\cite{senn3} where the QM/MM interactions are described at the MM level and therefore the electron density of the QM subsystem is not polarised by the MM subsystem. The use of pure QM and MM calculations makes this approach computationally advantageous, but prone to errors whenever the electron density of the QM subsystem is strongly polarised by the MM subsystem. The most popular coupling approach today is {\em electrostatic embedding}\cite{senn3} where the point charges of the MM subsystem are included in the one-electron Hamiltonian of the QM subsystem. This leads to a direct polarisation of the QM subsystem and usually a reasonable accuracy of its description. If electrostatic embedding is not enough to describe the coupling between the QM and MM subsystems, the final step is to use a {\em polarised embedding}\cite{senn3} approach, where the QM part is coupled to a polarisable force field, or to use hierarchical QM/QM layering.

The choice of QM method to be used used in a QM/MM simulation depends on the compromise between the accuracy needed for a given problem/system\cite{senn} and the computational resources available. Currently, methods based on density-functional theory\cite{parr,becke2} (DFT) are the most common choice, because of their favourable cost/accuracy ratio.\cite{senn, brunk, eichinger} Nevertheless, for some specific applications (e.g., involving photo-excited electronic states), the use of more accurate post-Hartree--Fock\cite{szabo} methods, such as M\o{}ller--Plesset perturbation theory,\cite{cremer} coupled-cluster theory,\cite{shavitt} or multiconfigurational methods, like complete active space self-consistent field (CAS-SCF),\cite{werner, shepard} is preferable.\cite{claeyssens, kongsted}

Having access to software offering a wide range of QM methods and different MM force fields would guarantee maximal flexibility in the choice of the most accurate and suitable QM/MM approach, as well as the possibility of consistent benchmarking for the least expensive approaches. 
To date, there exist a number of packages that are optimised toward specific QM or MM methods. Therefore, universal, flexible QM/MM could be more effectively achieved by coupling those specialised software than by rewriting a monolithic package dealing with all possible QM and MM implementations. MiMiC\cite{mimic, mimic2} is a recently developed framework for multiscale modeling in computational chemistry showcasing the wanted flexibility for easy yet efficient interfacing among different programs, with only small adaptions in the individual codes. So far MiMiC offers a coupling between the plane-wave DFT\cite{marx} program CPMD\cite{cpmd} and the widely used classical molecular dynamics (MD) program GROMACS.\cite{gromacs1,gromacs2} In that implementation, MiMiC uses CPMD as the main driver for the MD and for the description of the QM subsystem.

In this work, we present a wavefunction-based QM/MM implementation by coupling the CFOUR program package\cite{cfour} to the MiMiC framework. This allows the use of Hartree--Fock (HF), post-HF methods, like second-order M\o{}ller--Plesset (MP2) and truncated coupled-cluster methods (e.g., CCSD and CCSD(T)), and multiconfigurational methods within both ab-initio MD (AIMD) and electrostatic-embedding QM/MM-MD. The use of CPMD as the main MD driver also offers access to a multiple time step (MTS) algorithm\cite{liberatore} in which we can directly combine DFT with wavefunction-based methods. The MTS algorithm together with a long-range electrostatic coupling scheme\cite{laio} reduces the computational cost and makes QM/MM-MD simulations with high-accuracy QM methods feasible. 

The paper is organized as follows. In section \ref{theory} the implementation of the electrostatic interactions between the QM and MM subsystems in the quantum-chemical package CFOUR is described, before we discuss the coupling of CFOUR within the MiMiC framework. In section \ref{compdet} we outline the computational details of the test systems and the benchmark simulations. The results of these simulations are shown in section \ref{verification} where we demonstrate the computational stability and functionality of our implementation.

\section{Methods}\label{theory}

\subsection{QM/MM coupling Hamiltonian}

In the present QM/MM implementation, we use a full Hamiltonian electrostatic-coupling scheme, following ref. \citenum{laio}. The interaction Hamiltonian $\hH_{\mathrm{QM/MM}}$ is split into a bonded and a nonbonded part:
\be
\hH_{\mathrm{QM/MM}} = \hH_{\mathrm{bonded}} + \hH_{\mathrm{nonbonded}}.
\ee
The bonded part $\hH_{\mathrm{bonded}}$ is only nonzero if the QM/MM boundary cuts through covalent bonds, and treated at the molecular mechanics level. In this case the QM atoms at the boundary are replaced by monovalent pseudopotentials.\cite{lilienfeld,senn,mimic}
The non-bonded interactions consist of van der Waals interactions, which are described by the classical force field, and electrostatic interactions. For the latter, we adapted and implemented the electronic-coupling scheme developed by Laio et al.\cite{laio} in CFOUR. Because of the cost of the explicit treatment of electrostatic interactions, the MM atoms are divided into short- and long-range terms. The short-range contribution takes into account the explicit interactions between the nuclei and electrons of the QM subsystem and the point charges of the MM subsystem. In contrast, the interactions between the MM atoms belonging to the long-range region and the QM subsystem are calculated through a multipole expansion of the electrostatic potential from the QM electrons. This is possible because of the local character of the QM electron density and the distance to the long-range MM atoms:\cite{laio}
\be
\hH_{\mathrm{QM/MM}}^{\mathrm{el}} = \hH_{\mathrm{sr}} + \hH_{\mathrm{lr}}.
\ee
The short- and long-range regions are determined by a cut-off radius from the central QM part dividing the MM atoms into either region. This leads to the following short-range interaction Hamiltonian:
\be\label{hsr}
\hH_{\mathrm{sr}} =  - \sum_{i}^n \sum_{A}^{M_{\mathrm{sr}}} \frac{q_A}{|\bs{R}_A - \bs{r}_i|} +  \sum_{I}^N \sum_{A}^{M_{\mathrm{sr}}} \frac{q_A Z_I}{|\bs{R}_A - \bs{R}_I|}
\ee
where the sums run over all short-range MM atoms $M_{\mathrm{sr}}$, all electrons $n$, and all nuclei of the $N$ QM atoms. Thus, $q_A$ and $\bR_A$ are the point charge and coordinate of an MM atom, $\br_i$ is an electron coordinate, and $Z_I$ is the charge of a nucleus. This leads to an external potential that is included in the optimization of the wavefunction of the QM subsystem.

The forces on an MM atom and a QM nucleus due to the interactions between MM point charges and QM nuclei are given by the negative derivative of the second term in eq. \ref{hsr} with respect to the MM and QM coordinates, respectively:
\be\label{nnforce1}
\bs{F}_{IA}(A) = \sum_I^N \frac{q_A Z_I}{| \bR_A - \bR_I |^3} (\bR_A -\bR_I)
\ee
and
\be\label{nnforce2}
\bs{F}_{IA}(I) = -  \sum_A^{M_{\mathrm{sr}}} \frac{q_A Z_I}{| \bR_A - \bR_I |^3} (\bR_A - \bR_I).
\ee
The force on an MM atom due to its interaction with the QM electrons is given by the expectation value of the negative derivative of the first term in eq. \ref{hsr} with respect to the MM coordinates:
\be\label{mmsrforce}
\bs{F}_{iA}(A) = \sum_{\mu\nu} D_{\mu \nu} \left\langle \chi_\mu \bigg\vert \frac{q_A}{|\bR_A - \br|^3} (\bR_A- \br) \bigg\vert \chi_\nu \right\rangle
\ee
with the density matrix elements $D_{\mu\nu}$ and basis functions $\chi_\mu$ and $\chi_\nu$. The corresponding force on a QM nucleus is given by
\be\label{qmsrforce}
\bs{F}_{iA}(I) =  \sum_{A}^{M_{\mathrm{sr}}} \sum_{\mu\nu} D_{\mu \nu} \left[ \left\langle \frac{\delta \chi_\mu}{\delta \bR_I} \bigg\vert \frac{q_A}{|\bR_A - \br|} \bigg\vert \chi_\nu \right\rangle + \left\langle \chi_\mu \bigg\vert \frac{q_A}{|\bR_A - \br|} \bigg\vert \frac{\delta \chi_\nu}{\delta \bR_I} \right\rangle \right].
\ee
There are also implicit forces on the QM atoms due to the polarisation of the QM subsystem by the MM point charges and the consequent changes in the molecular orbitals.

In our implementation of the long-range interactions, we truncate the multipole expansion after the fourth order and only expand the electronic part of the electrostatic interactions. The interactions between QM nuclei and MM atoms in the long-range region are still calculated explicitly (compare eq. \ref{nnforce1} and \ref{nnforce2}). The multipole expansion (with the origin at zero) is given by
\be\label{multe}
\begin{split}
\sum_i \sum_A \frac{q_A}{|\bR_A - \br_i|} & = \sum_A q_A \Biggl[ C \frac{ 1}{|\bR_A|} + \sum_\alpha \mu^\alpha \frac{ R_A^\alpha}{|\bR_A|^3} + \frac{1}{2} \sum_{\alpha,\beta} \Theta^{\alpha \beta} \frac{R_A^\alpha R_A^\beta}{|\bR_A|^5} \\
 & + \frac{1}{6} \sum_{\alpha, \beta, \gamma} \Omega^{\alpha \beta \gamma} \frac{ R_A^\alpha R_A^\beta R_A^\gamma}{|\bR_A|^7} + \frac{1}{24} \sum_{\alpha, \beta, \gamma, \epsilon} \Phi^{\alpha \beta \gamma \epsilon} \frac{R_A^\alpha R_A^\beta R_A^\gamma R_A^\epsilon}{|\bR_A|^9}   \Biggr]
\end{split}
\ee
with charge $C$, dipole moment $\mu^\alpha$, quadrupole moment $\Theta^{\alpha \beta}$, octopole moment $\Omega^{\alpha \beta \gamma}$, and hexadecapole moment $\Phi^{\alpha \beta \gamma \epsilon}$ of the QM electron density. Here $\alpha$, $\beta$, etc.\ denote Cartesian components. Because of the use of atom-centered Gaussian-type orbital (GTO) basis functions in the QM part, we express the multipoles in an integral form. For the charge and the dipole moment the expressions have the following form:
\be
C = \sum_{\mu, \nu} D_{\mu\nu}\langle \chi_\mu | \chi_\nu \rangle \qquad \mathrm{and} \qquad 
\mu^\alpha = \sum_{\mu, \nu} D_{\mu\nu}\langle \chi_\mu | \alpha | \chi_\nu \rangle.
\ee
The forces on the MM atoms and QM nuclei are calculated by differentiation of eq. (\ref{multe}) with respect to the coordinates of the long-range MM atoms $R_A$ and with respect to the coordinates of the QM nuclei $R_I$ at which the basis functions are centered. The expressions for the derivatives are given in the appendix together with the explicit form of the quadrupole moment $\Theta^{\alpha \beta}$, the octopole moment $\Omega^{\alpha \beta \gamma}$, and the hexadecapole moment $\Psi^{\alpha \beta \gamma \epsilon}$.

The calculation of the long-range energy contribution and forces is computationally much cheaper than the explicit calculation for the short-range region. This is the case, because the calculation of the multipoles (and their derivatives) is decoupled from the sum over MM point charges whereas for the short-range region a one-electron integral must be calculated individually for every MM atom (see eq. \ref{hsr} and \ref{mmsrforce}). The resulting reduction in the computational cost is shown in section \ref{sectionlr}.

\subsection{Interface to the MiMiC framework}

\begin{figure}
\includegraphics[width=0.80\textwidth]{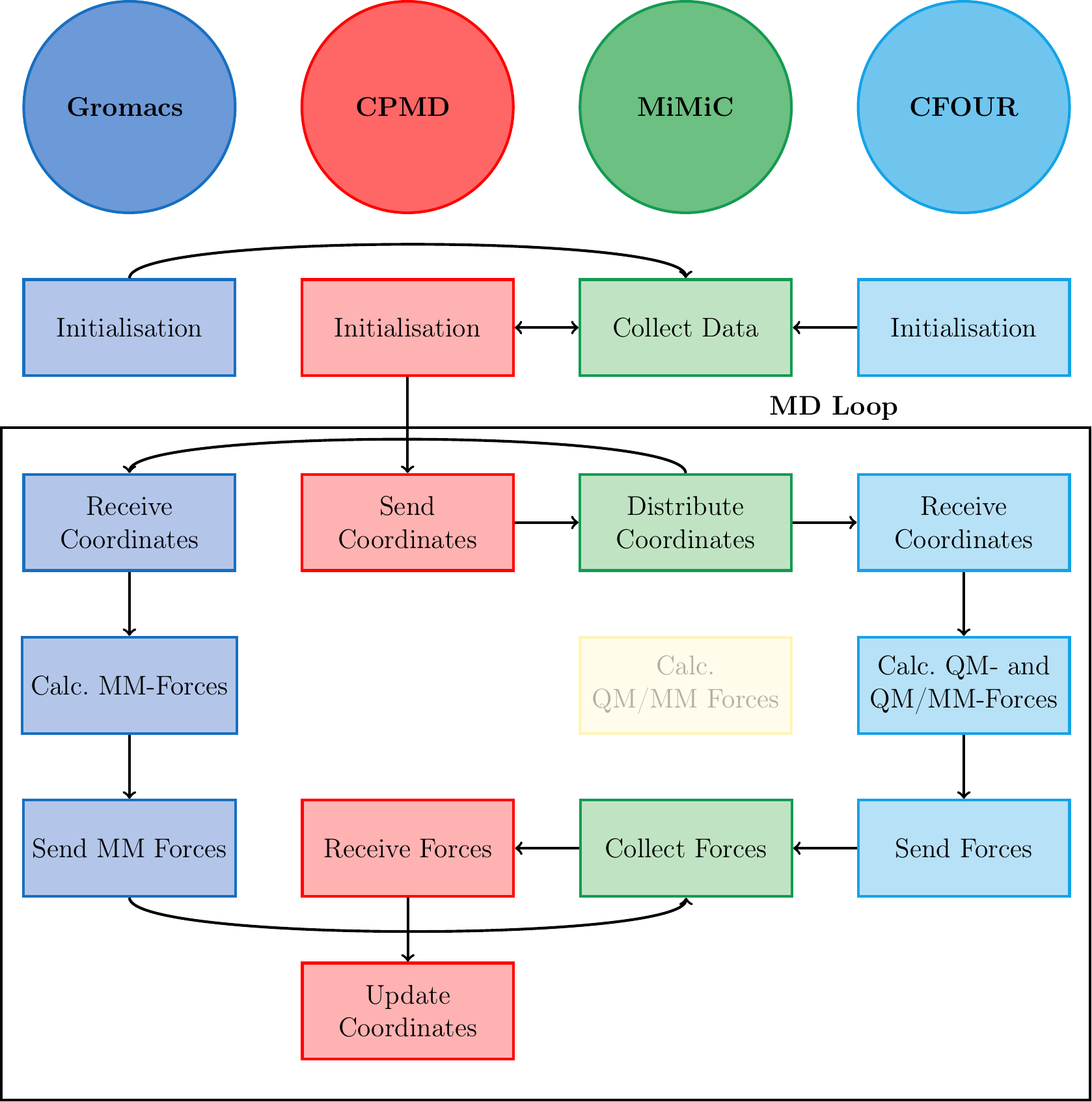}\centering
\caption{Scheme of the QM/MM-MD workflow using CFOUR (QM), GROMACS (MM) and CPMD (MD driver) via the MiMiC framework.}
\label{workflow}
\end{figure}

The CFOUR quantum-chemistry package was interfaced to the AIMD program CPMD and the classical MD package GROMACS through the MiMiC framework.\cite{mimic,mimic2} The MiMiC communication library is responsible for the exchange of all relevant information, like sending coordinates and charges of the MM atoms to CFOUR, and returning the calculated energy and forces necessary for each MD step via a message passing interface (MPI). The workflow of our QM/MM implementation is shown in figure (\ref{workflow}). 
In this implementation CPMD is the MD driver, GROMACS calculates the MM and van der Waals QM/MM energy and forces, and CFOUR calculates the QM energy and forces. In contrast to the MiMiC-based QM/MM implementation in ref. \citenum{mimic}, where the electrostatic QM/MM interactions are calculated by MiMiC (transparent yellow box in figure \ref{workflow}), in the present implementation these contributions are calculated by CFOUR. This is done because of the integral form of these terms, which numerically strongly depends on the functional form of basis functions used in the QM region. All three programs run independently on their own compute nodes while MiMiC manages the communication and data exchange between the programs.

\newpage

\subsection{Computational Details}\label{compdet}

To verify the QM/MM implementation, we used a small system consisting of one QM water molecule solvated by 1011 MM water molecules and a larger one consisting of one QM water molecule solvated by 12000 MM water molecules. The small and large systems are used for the analysis and validation of the short- and long-range coupling, respectively. All simulations were performed using the MiMiC framework coupling locally modified versions of GROMACS (version 2018) and CPMD (version 4.3), and a modified developer version of CFOUR. We also performed AIMD simulations on a single water molecule in vacuum that are used to compare with the solvated systems. For these AIMD simulations of water in vacuum, the starting geometries were obtained by a geometry optimization at the same level as the one used in the simulations, i.e., either HF, MP2, CCSD(T), or CAS-SCF(6,6) together with the cc-pVTZ\cite{dunning} basis set. In our case CAS-SCF(6,6) means an active space of 6 orbitals filled with 6 electrons for the CI calculation. 

The liquid water systems for QM/MM-MD simulations were built and preequilibrated by classical MM-MD simulations. We filled boxes of 3.164$^3$ and 7.184$^3$ nm with 1012 and 12001 water molecules, respectively. The structures were minimized with a steepest-descent scheme until the maximum force was lower than $1000$ kJ $\cdot$ (mol $\cdot$ nm)$^{-1}$. After the minimization, a 100 ps (2 fs time step) simulation was run in the NVT ensemble. The initial velocities were assigned from a Maxwell distribution at a temperature of 300 K that was maintained by a modified Berendsen thermostat\cite{berendsen} using one coupling group and a time constant of 0.1 ps. After that, we ran a 100 ps (2 fs time step) simulation in the NPT ensemble at 1.0 bar using a Parrinello--Rahman\cite{parrinello} barostat with a time constant of 2 ps. The temperature was again controlled by a modified Berendsen thermostat at 300 K. Finally, we performed a 1 ns (2 fs time step) simulation run in the NVT ensemble, again at 300 K and with the same parameters as before. For all the preequilibration steps, we used periodic boundary conditions, a cutoff of 1.0 nm for the short-range electrostatic and van der Waals interactions using the Verlet\cite{pall} scheme and all bonds involving H-atoms were constrained using the LINCS algorithm\cite{hesslincs} and the rigidity of the water molecules is ensured by the SETTLE algorithm.\cite{miyamoto} The long-range electrostatic interactions were calculated using the particle-mesh Ewald\cite{essmann} (PME) method. As integrator, we used the leap-frog scheme.\cite{hockney} In this way, we obtained equilibrated liquid water systems with 1012 water molecules in a cubic box with dimensions of 3.116 nm for the extended simple point charge (SPC/E)\cite{berendsen2} water model and one with dimensions of 3.129 nm for the three points (TIP3P)\cite{jorgensen2} water model, as well as a large liquid water system with 12001 SPC/E water molecules in a cubic box with dimensions of 7.119 nm.

The Born--Oppenheimer approach was used for all AIMD and QM/MM-MD simulations.\cite{marx} The simulation times were approximately 12.1 ps (50000 time steps of 10 a.u.). The AIMD simulations were initialized to a temperature of 1 K and the QM/MM-MD simulations on liquid water were initialized to a temperature of 300 K and none of the systems were coupled to a thermostat (NVE ensemble).

The MM subsystem was described by the TIP3P and the SPC/E water models. As in the preequilibration, periodic boundary conditions were applied for the MM subsystem, a cutoff of 1.5 nm was used for the short-range electrostatic, and van der Waals interactions together with a PME scheme for the long-range electrostatic interactions.

The QM subsystem was described by either plane-wave (PW) based DFT (PW-DFT) or a wavefunction-based method. In case of PW-DFT, we used the BLYP exchange--correlation functional\cite{becke,lee} with Troullier--Martins norm-conserving pseudopotentials.\cite{troullier} We used isolated system conditions for the QM subsystem and the Tuckerman \& Martyna\cite{martyna} scheme to solve Poisson's equation. The cell size was 30 a.u.\ and we used a PW cutoff of 100 Ry. For the SCF optimization, a convergence criterion of $10^{-5}$ a.u.\ for the gradient of the orbitals was used.
In case of wavefunction-based methods, the correlation consistent cc-pVTZ basis\cite{dunning} was used in all cases. The SCF convergence criterion was $10^{-9}$ a.u.\ for the density matrix and, in case of the CCSD(T) calculations, convergence criteria of $10^{-7}$ a.u.\ were used for the CC- and lambda equations.

All electrostatic QM/MM interactions on the system with 1012 water molecules were computed without periodic conditions using the short-range coupling for the entire system. For the system with 12001 water molecules, both short- and long-range electrostatic interactions were used with different cutoff distances.
The dipole moment of the QM system was calculated on the fly at every step of the simulations. For the analysis of the systems and evaluation of properties, the last 30000 steps of the simulations were used. 

\clearpage

\section{Results}

\subsection{Validation}\label{verification}

\subsubsection{Energy conservation}\label{energycons}

\begin{figure}[ht!]\centering
     \includegraphics[width=\textwidth]{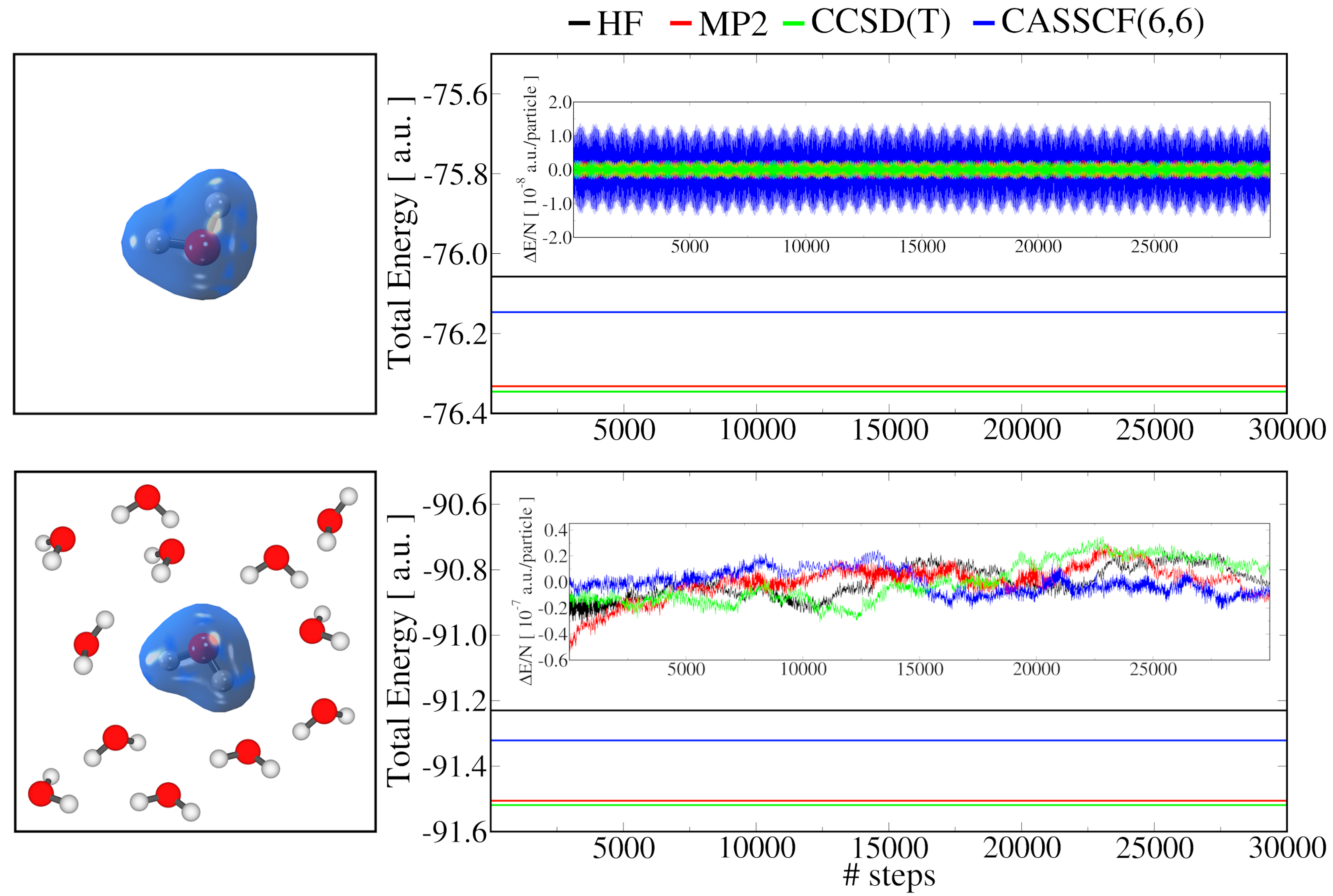}
    \caption{Total energy of a single water molecule (top part) and of the QM/MM system with 1012 water molecules (bottom part) simulated within the NVE ensemble using HF, MP2, CCSD(T), and CAS-SCF(6,6) with the cc-pVTZ basis to model the QM water molecule and SPC/E for all MM water molecules. The insets show the fluctuation of the total energy per atom for the respective system.}
    \label{nrg_fig}
\end{figure}

Conservation of the energy during a simulation is an important indicator of a stable AIMD or QM/MM-MD implementation and guarantees that the interface is suitable to sample a thermodynamic ensemble. Therefore, we investigated the profile of the energy as well as the deviation from the average energy per particle during a NVE simulation of one water molecule in the gas phase. The results for the AIMD simulations using HF, MP2, CCSD(T), and CAS-SCF(6,6) are shown in the top part of figure (\ref{nrg_fig}).
The energy deviation is calculated by $ \Delta E(t) = E(t) - \langle E \rangle $. For all QM methods, the energy fluctuation is very small and there is no visible drift in the energy. For CAS-SCF(6,6) comparatively larger fluctuations for the single water molecule are observed, but with a standard deviation of $2.4 \cdot 10^{-8}$ a.u.\ they anyway remain within negligible values.

We also examined the energy fluctuation for the QM/MM-MD simulations using HF, MP2, CCSD(T), and CAS-SCF(6,6) for the QM water molecule solvated by 1011 classical SPC/E water molecules. The results are shown in the bottom part of figure (\ref{nrg_fig}). Also in this case, the energy fluctuations are very small and no significant drift in the energy is seen. The standard deviation of the energy per particle is $3.7 \cdot 10^{-5}$ a.u.\ for HF, $4.0 \cdot 10^{-5}$ a.u.\ for MP2, $4.9 \cdot 10^{-5}$ a.u.\ for CCSD(T), and $2.7 \cdot 10^{-5}$ a.u.\ for CAS-SCF(6,6). These results show that our AIMD and QM/MM-MD implementation enables stable simulations using various wavefunction-based QM methods.

\subsubsection{Polarisation of the QM region}\label{dipole}

An important aspect in QM/MM approaches is the description of the effect of the MM environment on the structure and polarisation of the QM part.\cite{cascella} To test that, we investigated two properties as indicators of the quality of the QM/MM coupling, namely, the dipole moment of one QM water molecule surrounded by 1011 SPC/E water molecules, and its radial distribution function\cite{soper} (RDF).

In contrast to classical MM-MD simulations, QM/MM-MD (and AIMD) simulations give detailed information about the electronic structure of the solute. Thus we can analyse bonding and electronic properties such as the dipole moment. 

\begin{table}[ht]\centering\scriptsize
\begin{threeparttable}
\caption{Dipole Moment of a QM Water Molecule\textsuperscript{\emph{a}}}
\label{tab_dipole}
\begin{tabular}{lllllll}
\toprule \toprule
 method & HF & MP2 & CCSD(T) & CAS(6,6) & BLYP & exp. \\
 \midrule
single-point		 & 	1.988		&	1.936	&	1.916	&	1.935	&	1.814	&	1.847 $\pm$ 0.001 \cite{clough}		\\
AIMD (in vacuo) 		&	1.988 $\pm$ 0.004	&1.936 $\pm$ 0.004	&1.916 $\pm$ 0.004	&	1.909 $\pm$ 0.031	&	1.846 $\pm$ 0.005 &	\\	
QM/MM (SPC/E)	&	2.694 $\pm$ 0.160	&2.742 $\pm$ 0.145&2.765 $\pm$ 0.166	&	2.624 $\pm$ 0.177&	2.720 $\pm$ 0.169	&	2.9 $\pm$ 0.6\cite{badyal}	\\
QM/MM (TIP3P) &	2.732  $\pm$ 0.146		&2.692  $\pm$ 0.166	&	2.632  $\pm$ 0.168		&	2.755 $\pm$ 0.157	& 2.699	$\pm$ 0.200	& \\	\bottomrule  \bottomrule 
\end{tabular}
\textsuperscript{\emph{a}} The dipole moment (in Debye) is obtained from single-point QM calculations as well as AIMD (at 1~K) and QM/MM-MD simulations (at 300~K) using the cc-pVTZ basis for the wavefunction-based QM methods. Standard deviations are given for the dipoles obtained from MD simulations.
\end{threeparttable}
\end{table}

It is well known that the polarisation of the electronic structure of a water molecule in liquid water leads to an increase of the dipole moment compared to the gas phase.\cite{silvestrelli,grubskaya,liu,scipioni,badyal}
To get a reference for the dipole moment of an unpolarised water molecule, we performed AIMD simulations of a single water molecule as well as single-point calculations on the geometry-optimized structure. For a polarised water molecule, we calculated the average dipole moment of the QM water molecule from the QM/MM-MD trajectories of the small liquid water system. The results are shown in table (\ref{tab_dipole}). The single-point and average dipole moments of gaseous water are the same (minor deviation for CAS-SCF and DFT), as expected, and it is again a good indication that the AIMD implementation works well. All gas-phase values agree well with the experimental value, within the approximation associated to their respective level of theory.

The dipole moment of a water molecule in the liquid phase obtained from the QM/MM-MD simulations are, for every QM method, much larger than the gas-phase dipole moment. This is mainly a result of the polarisation of the electronic structure due to the environment. The dipole moments of the two SPC/E and TIP3P classical water models are only slightly different ($2.351$ D for SPC/E\cite{berendsen2} and $2.344$ D for TIP3P\cite{jorgensen2}), and produce a similar polarisation of the QM water (Table~\ref{tab_dipole}). For both classical models, the standard deviation of the dipole moment of the QM water is quite large. The significant shift in the average value of the water dipole from gas to liquid phase, and the large fluctuations of the dipole moment during the simulation indicate that our implementation correctly captures the very strong coupling between the electronic density and the  surrounding mobile classical charges. 
The obtained values are somewhat smaller than the experimental value but still well within the uncertainty and also match well with other theoretical studies of the dipole moment of liquid water.\cite{silvestrelli,grubskaya,liu,scipioni,badyal} 
Since our QM/MM implementation only considers the polarisation of the QM subsystem due to the MM subsystem using water models with fixed dipole moments that are in general smaller than the experimental values, it is plausible to find that also the dipole moment of the QM water is smaller than the experimental one. In general, quantitatively better results 
 for liquid water require that mutual polarisation and other effects between the central and the surrounding waters are taken into account. These include (among others) charge transfer and better descriptions of Pauli repulsion and dispersion. Therefore, either larger QM regions or a more advanced water model would be needed, for example polarisable variants like the MB-pol\cite{reddy} force-field, which is constructed including many-body terms calculated at the CCSD(T)/CBS level of theory.

\begin{figure}[hbt]
\centering
\includegraphics[width=0.95\textwidth]{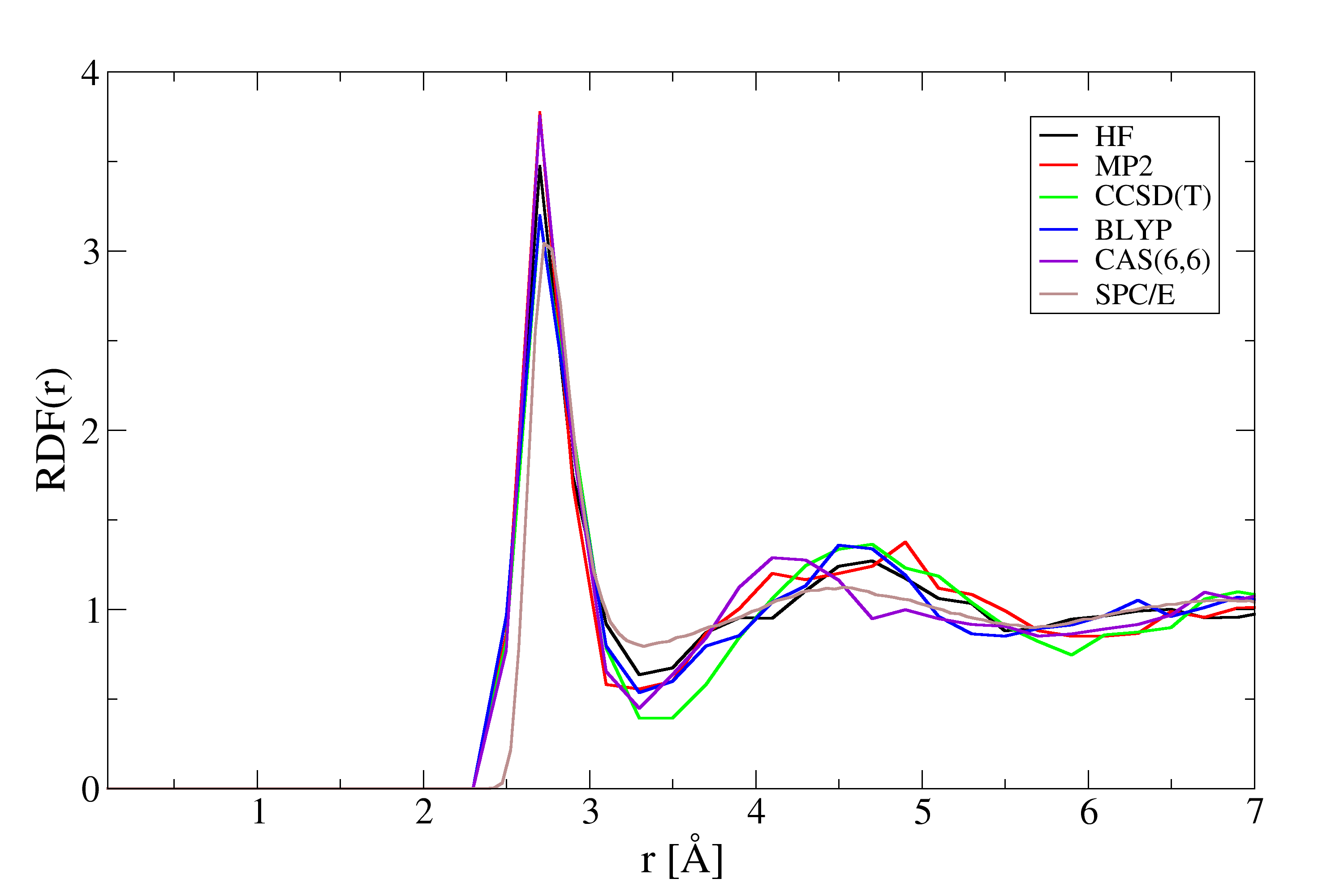}
\caption{Oxygen--oxygen radial distribution function (RDF) between the QM and MM oxygen atoms at 300~K from QM/MM-MD simulations within the NVE ensemble using the SPC/E water model and from a pure MM-MD simulation with the SPC/E water model. The cc-pVTZ basis was used for the wavefunction-based QM methods.}
\label{rdf_spce}
\end{figure}

The structure of the QM water and its solvation shell was monitored by calculation of the RDF between the QM and MM oxygens. 
The results for the different QM methods and for a pure SPC/E MM-MD simulation are shown in figure (\ref{rdf_spce}). For all QM methods, the distances of the first maximum is in very good agreement and for the second maximum still in good agreement with the experimental values\cite{soper,badyal} (see table (\ref{tab_rdf})). The overall shape of the RDF looks as expected, apart for some noisy features due to relatively poor sampling.
This agreement shows that the environment of the QM water molecule and especially the hydrogen-bonded neighbours are described well with our QM/MM implementation. Compared to the pure SPC/E model, the QM water appears to rigidify the local structure of its first neighbours, with a higher first peak, and a lower first minimum, regardless of the QM level of theory, in agreement with what has been observed in past studies on QM water models\cite{lin,lin2,miceli}.

\begin{table}[ht]\centering
\begin{threeparttable}
\caption{First and Second Peak of the Oxygen--Oxygen RDF\textsuperscript{\emph{a}}}
\label{tab_rdf}
\begin{tabular}{lccccccc}
\toprule \toprule
maxima & HF & MP2 & CCSD(T) & CAS(6,6) & BLYP & SPC/E & exp.\cite{soper} \\
 \midrule
1. $r_{\mathrm{OO}}$ & 2.82 & 2.80 & 2.81 & 2.80 & 2.81 & 2.89 & 2.8 \\ 
2. $r_{\mathrm{OO}}$ & 4.55 & 4.80 & 4.75 & 4.24 & 4.54 & 4.61 & 4.5  \\ 
\bottomrule  \bottomrule 
\end{tabular}
\textsuperscript{\emph{a}} The distances (in \AA{}) for the first and second peak of the oxygen--oxygen radial distribution functions (RDFs) between the QM and MM oxygen atoms at 300~K are obtained from QM/MM-MD simulations within the NVE ensemble using the SPC/E water model. The cc-pVTZ basis was used for the wavefunction-based QM methods.
\end{threeparttable}
\end{table}

\subsubsection{Vibrational frequencies}

To test the quality and reliability of the dynamical properties obtained by our implementation, we calculated vibrational frequencies from MD  trajectories,\cite{thaunay,thomas,wang,vitale,silvestrelli2,praprotnik,liu,giovanni} 
obtaining them from the Fourier transform of the dipole-moment autocorrelation function.\cite{thaunay,wang,thomas,vitale}

\begin{figure}[ht!]
    \centering
        \begin{subfigure}[h]{1.0\textwidth}
            \centering
            \includegraphics[width=\textwidth]{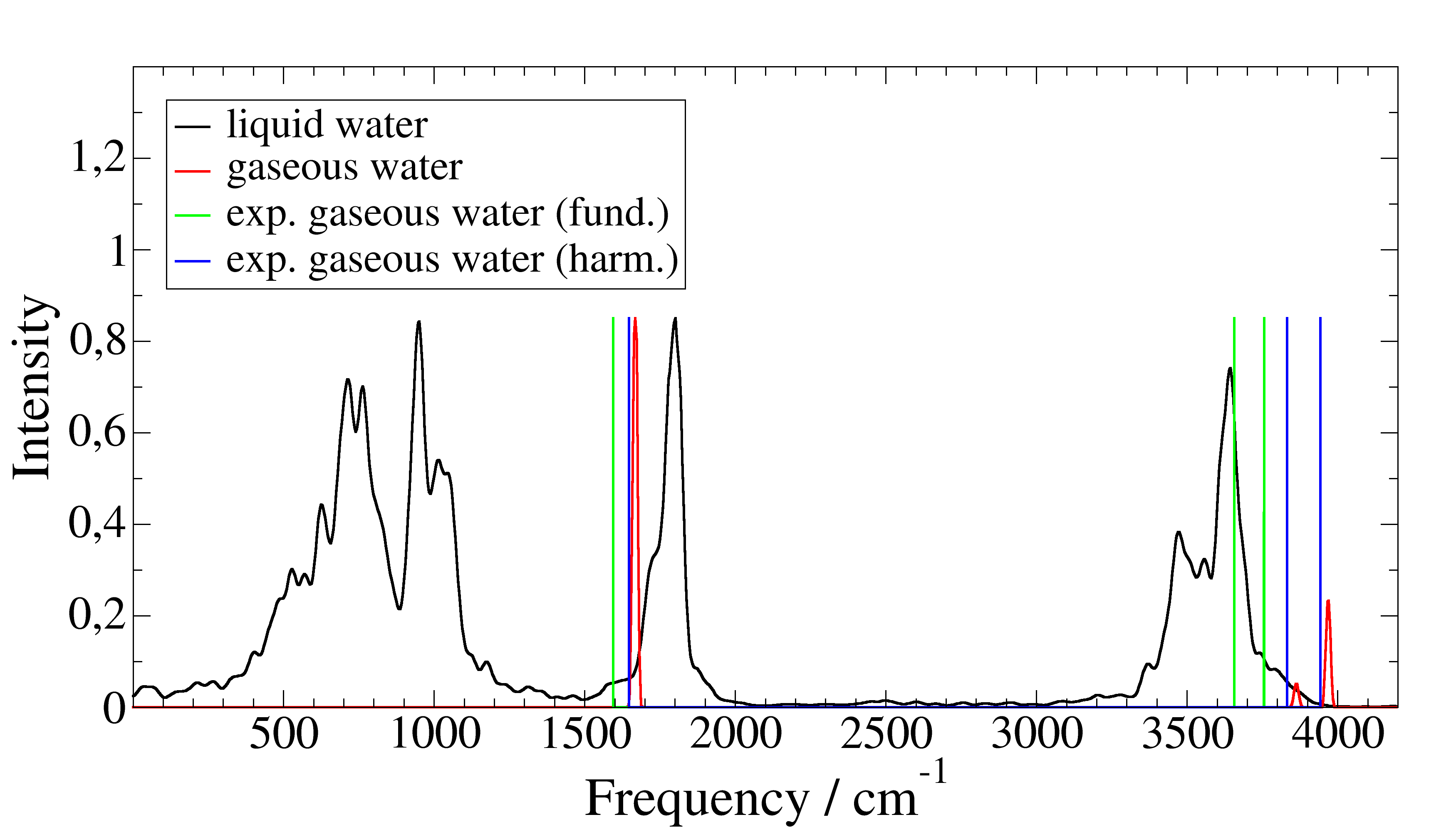}
            \caption{ }
            \label{ir_spce}
        \end{subfigure}\vspace{0.5cm}
        \begin{subfigure}[h]{1.0\textwidth}
            \centering\footnotesize
            \begin{tabular}{lcccccc}
                \toprule \toprule
                     Mode & SP & AIMD & QM/MM & \makecell{exp. gas \\ (harm.)\cite{toukan}} & \makecell{exp. gas \\ (fund.)\cite{lemus}} & \makecell{exp. liquid \\ (fund.)\cite{walrafen}}  \\
                 \midrule
                 bend & 	1667.3		&	1667.6	&	1772.0	&	1648 	&	1595		&	1640 \\
                 asym. stretch & 3858.3	&	3863.1	&	3477.5	&	3832		&	3657		&  3450	 \\
                 sym. stretch & 	3964.0	&	3969.1	&	3643.2	&	3943		&	3756		& 3615	 \\
                \bottomrule  \bottomrule 
            \end{tabular}
            \caption{ }
            \label{tab_ir}
        \end{subfigure}
    \caption{IR spectra (a) and frequencies (b) of gaseous and liquid water obtained from single-point (SP) calculations as well as AIMD (at 1~K) and QM/MM-MD simulations (at 300~K within the NVE ensemble. The CCSD(T)/cc-pVTZ level of theory was used throughout and the SPC/E water model was used for QM/MM. Experimental harmonic and fundamental frequencies are included for comparison. Frequencies are given in cm$^{-1}$.}
    \label{vib_2}
\end{figure}

Figure (\ref{ir_spce}) reports the IR spectrum of gaseous and liquid water using the same trajectories as in section \ref{dipole}. For the gaseous water molecule, we also calculated single-point harmonic vibrational frequencies at the CCSD(T)/cc-pVTZ level to compare to the AIMD results and to experiment, shown in figure (\ref{tab_ir}). The frequencies derived from AIMD simulations agree well with the frequencies obtained from the single-point calculation and also with the harmonic experimental values. Because AIMD simulations ran at very low temperature (1~K) we expected negligible temperature effects and no appearance of anharmonic deviations. For the solvated water system dynamical environment effects are essential for the position and the shape of the IR peaks. The IR spectrum obtained from the QM/MM-MD simulation has the typical broad band around $3500-3700$ cm$^{-1}$ that is caused by hydrogen-bond interactions interfering with the stretching modes. The obtained frequencies for the two stretching modes (3477.5 cm$^{-1}$ and 3643.2 cm$^{-1}$) are in a reasonable agreement with the experimental values (3450 cm$^{-1}$ and 3615 cm$^{-1}$). The position of the bending mode (1772.0 cm$^{-1}$) is at a higher frequency than the experimental value (1640 cm$^{-1}$). The comparatively large blue shift of the bending mode indicates the relatively strong influence of the water model. The solvent induced blue shift of the bending mode and the red shift of the stretching modes are qualitatively well reproduced which is another indication of a correct implementation of the QM/MM interface.

\subsection{Speeding up simulations}\label{speedup}
The use of highly accurate quantum-chemical methods in MD simulations is computationally very expensive. Therefore it is important to find ways to reduce the cost and to increase the efficiency of the implementation. Here we present two methods implemented in the CFOUR interface to MiMiC to speed-up simulations.

\subsubsection{Long-range interactions}\label{sectionlr}

Hierarchical treatment of the long-range electrostatic coupling is an excellent strategy to reduce the time per MD step in the simulation.\cite{laio} In the present implementation, the long-range electrostatic interactions are implemented in the CFOUR code, because of the integral form of the multipoles and their derivatives and the use of atom-centered GTO basis functions in the QM part. To investigate the effectiveness of the new implementation in reducing the computational cost, as well as the dependence of the numerical accuracy on the number of MM atoms in the short-range region, we performed single-point QM/MM calculations and QM/MM-MD simulations of one QM water molecule surrounded by 12000 SPC/E water molecules, which is a typical system size in studies of enzymatic reactions, using different cutoff distances $d_{\mathrm{cut}}$. The accuracy of the QM and QM/MM energy and forces as well as the dipole moment of the QM subsystem is tested by single-point calculations. The MD simulations show the stability of the long-range implementation and of the obtained properties. We compared the obtained values with calculations where all MM atoms are in the short-range region.

\begin{figure}[ht!]\centering
     \includegraphics[width=1.0\textwidth]{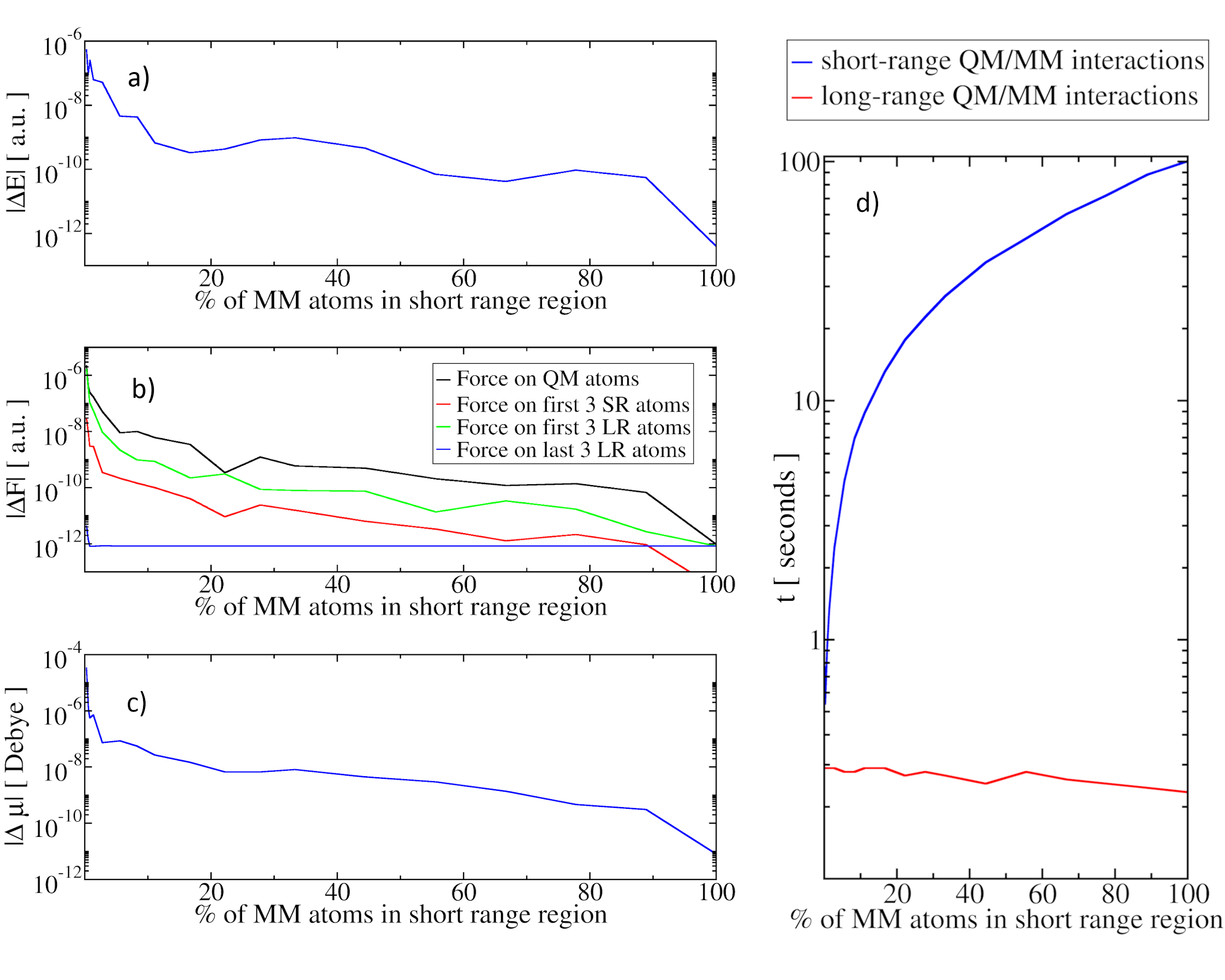}
    \caption{Convergence of the a) energy, b) forces, and c) dipole moment, and the d) absolute time with respect to the number of MM atoms in the short-range region for the calculation of the short- and long-range electrostatic QM/MM interactions in a single-point HF/cc-pVTZ calculation on one snapshot of a water system with 12001 water molecules. The reference is a calculation with all MM atoms in the short-range region. The energy in a) is the full QM energy plus the electrostatic QM/MM interaction energy. In b) the mean absolute error of the AIMD and electrostatic QM/MM forces on four different groups of atoms is shown (the three QM atoms, the three short-range MM atoms nearest the center of mass of the QM subsystem, the three nearest long-range MM atoms, and the three farthest long-range MM atoms). Plot c) shows the dipole moment of the QM subsystem and d) shows the relative time for the calculation of the QM/MM interactions (energy and forces).}
    \label{sp_lr}
\end{figure}

Figure (\ref{sp_lr}) shows the results of the single-point calculations. The convergence criterion for the density matrix in the solution of the HF equations for the single-point calculations was $10^{-12}$ a.u.\ which is much tighter than what we usually use in QM/MM-MD simulations ($10^{-7}$ - $10^{-9}$ a.u.).
It is seen that the errors are small for all numbers of MM atoms in the short-range region except of very low numbers (less than 5-10 $\%$ short-range atoms). We achieve an accuracy of about $10^{-9}$~a.u.\ for the energy by including only $10$ $\%$ of the MM atoms in the short-range region. This error is negligible compared to the fluctuations of the energy because of the time discretization error and thermal energy fluctuations at 300~K (compare \ref{energycons}).  Including more atoms in the short-range region improves the energy only slightly. The full convergence to the short-range value is only achieved if nearly all atoms are included.
This behaviour is also seen for the forces on the different groups of atoms, except for the three long-range atoms that are furthest from the QM subsystem, where the force is already exact even for very few short-range atoms. For the dipole moment the accuracy is about $10^{-8}$~D with $10$ $\%$ short-range atoms. This accuracy is sufficient for QM/MM-MD simulations where we usually use convergence criteria of similar magnitude. Thus, the number of MM atoms in the short-range region that are needed to guarantee stable simulations with the same accuracy as pure short-range simulations is about $10$ $\%$ for this system ($3600$ atoms). Nevertheless, this number can be different for other systems, e.g., with a larger or differently shaped QM subsystem, and should therefore be checked before performing a simulation. Figure (\ref{sp_lr}d) shows the reduction in computational cost for the electrostatic QM/MM interactions. 
As expected, the time for the short-range interactions increases linearly with the number of MM atoms in the short-range region. The time for the calculation of the long-range interactions is negligible in comparison to the time spent calculating the short-range interactions and increases only slightly with increasing number of long-range MM atoms. For example, the calculation of the short-range QM/MM interactions with all MM atoms in the short-range region took 102 seconds. The calculation of the QM/MM interactions with only $4000$ atoms ($11$~$\%$) short-range MM atoms took $9$ seconds, thereof only $0.3$ seconds were needed for the calculation of the long-range QM/MM interactions. Therefore, as expected, also in the present implementation the use of long-range interactions drastically reduces the computational cost without any significant loss of accuracy.

\begin{figure}[h]
\includegraphics[width=1.0\textwidth]{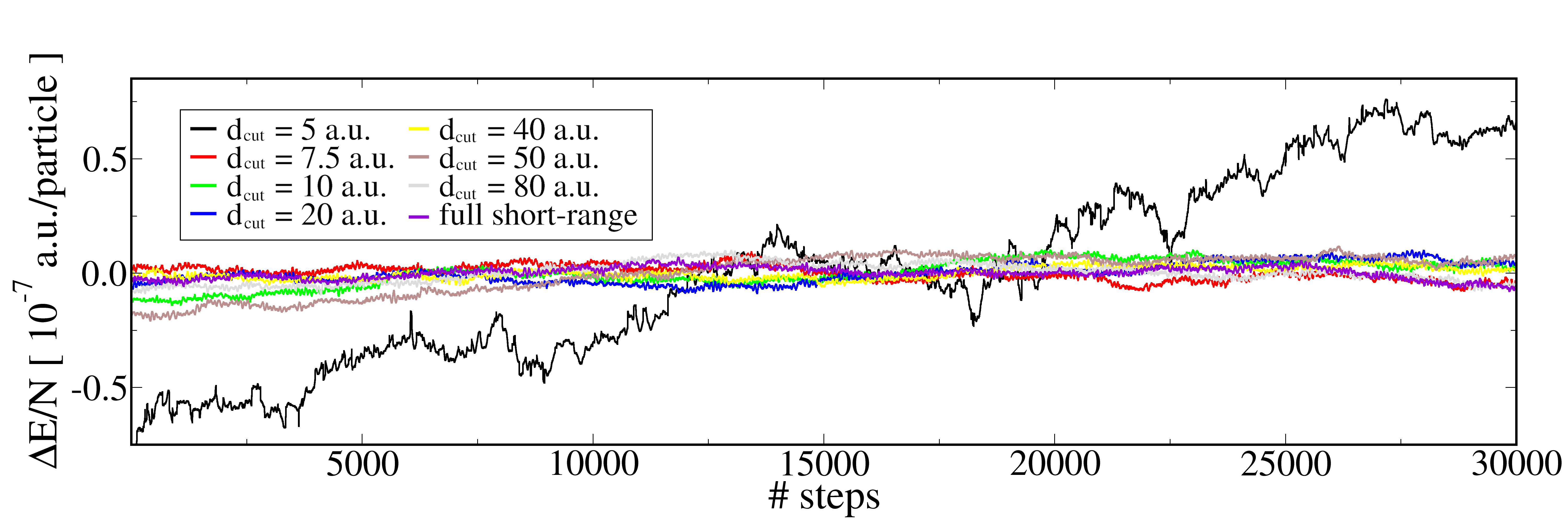}\centering
\caption{Fluctuation of the energy per particle for different cutoff distances for the long-range coupling. The system consisting of 12001 water molecules was simulated within the NVE ensemble at the HF/cc-pVTZ level of theory and using the SPC/E water model.}
\label{sim_lr}
\end{figure}

The fluctuation of the energy during a QM/MM-MD simulation with different cutoff distances is shown in figure (\ref{sim_lr}). There is a drift of the energy for the $5$ a.u.\ cutoff distance. No drift is observed for all other cutoff distances and the fluctuations are very small and of the same magnitude as for the full short-range coupling case. The standard deviation of the fluctuations of the energy per particle is between $3.0 \cdot 10^{-4}$ and $7.0 \cdot 10^{-5}$ a.u.\ for the different cutoff distances (excluding the $5$ a.u.\ cutoff). This shows that a cutoff distance of $7.5$ a.u.\ (which is only 35-50 MM atoms in the short-range region) is enough for a stable simulation. However, this would affect the accuracy compared to a full short-range treatment as discussed earlier. For comparison, a cutoff distance of $35-40$ a.u.\ corresponds to $10$ $\%$ MM atoms in the short-range region.

Table (\ref{lr_dip}) shows the average dipole moment and associated standard deviation for different cutoff distances. The dipole moments are all in the same range and within the standard deviation of the full short-range coupling case. They should converge to the same value when simulating long enough considering the precision given in the table and the accuracy that can be obtained (see figure (\ref{sp_lr}c)). This shows that even small cutoff distance like $10$ a.u., where only 60-90 MM atoms are in the short-range region, can be sufficient to adequately describe the polarisation of the electronic structure of the QM region. 

\begin{table}[ht]\centering
\begin{threeparttable}
\caption{Dipole Moment of the QM Water Molecule\textsuperscript{\emph{a}}}
\label{lr_dip}
\begin{tabular}{lrrrrrrrrr}
\toprule \toprule
 $d_{\mathrm{cut}}$  & 5 & 7.5 & 10 & 20 & 30 & 40 & 50 & 80 & only sr \\
 \midrule
 $\mu$      & 2.811   & 2.639  & 2.753  & 2.637  & 2.748   &  2.692  & 2.706  & 2.719  &  2.711  \\
$\sigma$    &  0.147  & 0.150  & 0.150  & 0.144  &  0.181  &  0.185  & 0.154  & 0.171  &  0.178   \\
\bottomrule  \bottomrule 
\end{tabular}
\textsuperscript{\emph{a}} The average dipole moment and associated standard deviation (in D) of the QM water molecule in the system with 12001 water molecules is obtained from a QM/MM-MD simulation at the HF/cc-pVTZ level of theory and using the SPC/E water model. The values are obtained for different cutoff distances for the long-range coupling and compared with the full short-range case.
\end{threeparttable}
\end{table}

\subsubsection{QM/QM multiple time step dynamics}\label{sectionmts}

Another method to speed-up the simulation is the use of an MTS algorithm.\cite{martyna,tuckerman,liberatore} This reduces the number of calculations needed at the high-accuracy quantum-chemical level. Because the maximum time step $\delta t$ used in the integration of the equation of motion is limited, MTS algorithms separate different degrees of freedom which can be integrated at different rates. The MTS algorithm implemented in CPMD\cite{liberatore} is an adaptation of the rRESPA\cite{tuckerman2} scheme and enables the use of a combination of quantum-chemical methods within a simulation. Here we explore the combination of BLYP and CCSD(T). Instead of using CCSD(T) in every step of the simulation, we use BLYP as the reference method, while CCSD(T) is used only at larger time steps as a correction. 

\begin{figure}[hbt]
\includegraphics[width=1.0\textwidth]{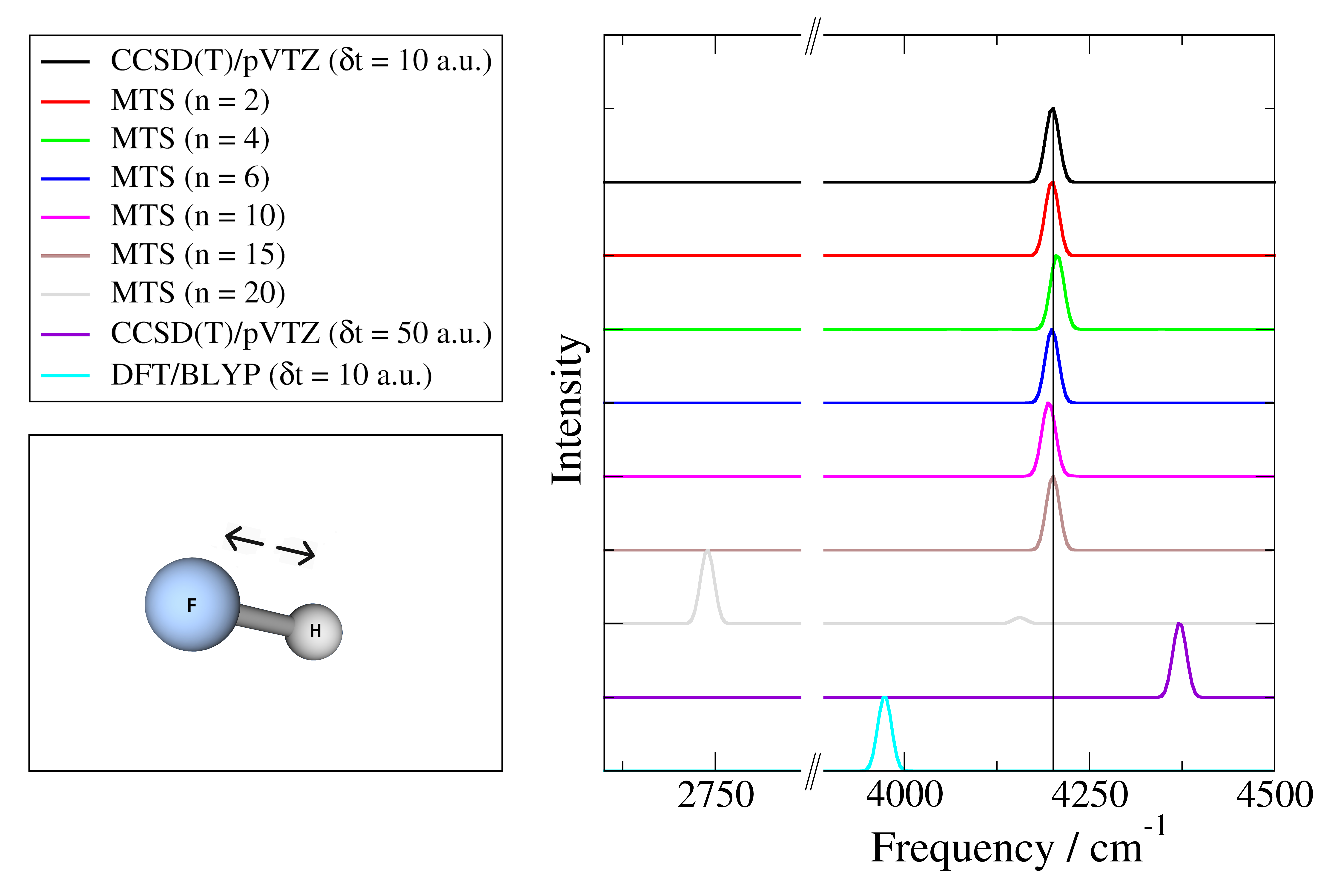}\centering
\caption{Vibrational spectra of hydrogen fluoride in vacuum obtained from AIMD simulations at the combined CCSD(T)/cc-pVTZ and BLYP/PW levels of theory with CCSD(T)/cc-pVTZ calculations only every $n$th step where $n = 2, 4, 6, 10, 15,$ or $20$. Frequencies are given in cm$^{-1}$.}
\label{hf_mts}
\end{figure}

We performed AIMD simulations on hydrogen fluoride, which has only one vibrational mode so that no other effects can interfere, to investigate the use of the MTS scheme within our implementation. To obtain reference data, we first performed simulations with a standard time step of 10 a.u.\ ($\approx$ 0.2 fs) at the CCSD(T)/cc-pVTZ or BLYP/PW levels of theory. Then, we performed simulations at the BLYP/PW level of theory with the use of CCSD(T)/cc-pVTZ calculations every $n$th time step (the MTS factor $n$ being $2, 4, 6, 10, 15,$ or $20$) as well as a simulation at the CCSD(T)/cc-pVTZ level of theory with a time step of 50 a.u.\ ($\approx$ 1.2 fs) to demonstrate that a CC simulation with an increased time step without any reference forces at lower level does not reproduce the correct value. The obtained spectra are given in figure (\ref{hf_mts}) and the energy fluctuations for the MTS simulations are given in the appendix.

The frequency obtained from BLYP/PW and CCSD(T)/cc-pVTZ simulations using the 10 a.u.\ time step differ substantially as expected. All BLYP/CCSD(T) MTS-MD simulations, except for the one with the largest MTS factor of 20, reasonably reproduce the pure CCSD(T) frequency. The simulation with an MTS factor of 4 has a deviation of $+6$ cm$^{-1}$ and the simulation with the MTS factor of 10 has a deviation of $-4$ cm$^{-1}$. The other simulations have deviations smaller than $\pm1$ cm$^{-1}$. However, the simulation with the largest MTS factor does not give reasonable results. Also the CCSD(T)/cc-pVTZ simulation with a time step of 50 a.u.\ does not reproduce the correct frequency because the time step is to large to describe the vibrational mode of hydrogen fluoride. Remarkably, the BLYP/CCSD(T) MTS-MD simulation can reproduce the pure CCSD(T) frequency even though a CCSD(T) calculation is only performed every 15th step (150 a.u. or $\approx$ 3.6 fs). All MTS-MD simulations reproduce the coupled-cluster average bond length of HF of $0.916 \pm 0.002~\AA$. Even the simulation with $n=20$, with an average bond length of $0.9120 \pm 0.03~\AA$,  reproduces reasonably well the CC value.

\begin{figure}[hbt]
\includegraphics[width=1.0\textwidth]{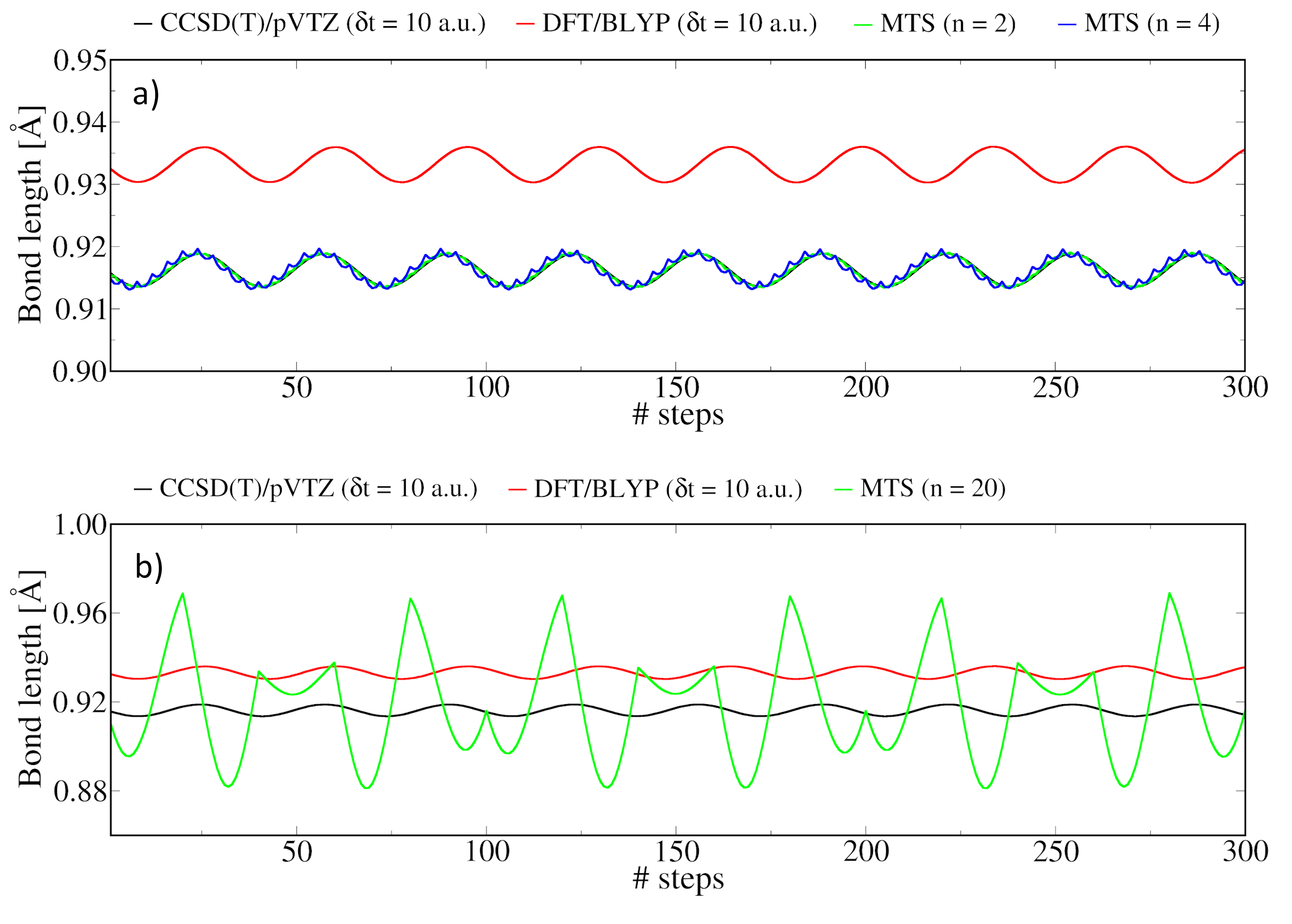}\centering
\caption{Fluctuation of the bond length of hydrogen fluoride in vacuum obtained from AIMD simulations at the combined CCSD(T)/cc-pVTZ and BLYP/PW levels of theory.}
\label{mts_bond}
\end{figure}

Figure (\ref{mts_bond}) shows the fluctuation of the bond length of hydrogen fluoride. Panel a) shows that in the MTS simulation a correction due to the CC calculations is introduced. These corrections appear in the dynamics as fast oscillations along the CC trajectory with a period corresponding to that of the correction rate. The oscillation of the correction and the physical oscillation remain well separated until the MTS factors are small enough, but for an MTS factor of $n=20$ the interval for the corrections is too big to reproduce the CC trajectory. The vibration introduced by the correction steps can also be seen in the obtained vibrational spectra. The MTS factor $n=2$ corresponds to a very high correction frequency of 68900 cm$^{-1}$. The correction frequency for the MTS factor of $n=20$ is at 6890 cm$^{-1}$ which is rather close to the hydrogen fluoride vibration. The critical value for the MTS correction to well reproduce the  CCSD(T) vibrational frequency is n=15 (with a correction frequency of 9187 cm$^{-1}$). Considering that one period of the hydrogen fluoride vibration takes around 33 time steps, we deduce that the MTS factor should be at least smaller than half of the period of the fastest vibrational mode that needs to be accurately reproduced. This is in agreement with the literature describing limitations of the MTS procedure and possible solutions to the resonance problem.\cite{liberatore, minary, leimkuhler, morrone} 

Another limitation to the maximum speedup of the method described by Liberatore et al.\cite{liberatore} is the increased number of SCF cycles needed in the wavefunction optimization in the correction steps for larger MTS factors. In our case this is negligible. The relative speedup for the simulation with the MTS factor $n=20$ is 19.78. This has two reasons: The number of the SCF cycles increases only slightly from 9/10 cycles in the case of a pure CCSD(T) simulation to 10/11 in the case of the simulation with MTS factor $n=20$. Furthermore, the time needed for the optimization of the wavefunction is negligible compared to the time needed for solving the CC equations which is independent of the initial guess of the wavefunction.    

This means that a substantial speedup of the simulation can be achieved with little to no loss in accuracy using MTS. However, the magnitude of the speedup can differ for bigger and more complicated systems and according to the different properties for which a higher-level correction is required.

\section{Summary and outlook}

We presented a wavefunction-based electrostatic-embedding QM/MM implementation enabled by the coupling of the CFOUR program package to the MiMiC framework. This allows the use of various QM methods like HF, MP2, CC, and CAS-SCF in AIMD and QM/MM-MD simulations. The implementation features an efficient long-range electrostatic coupling akin to the one in MiMiC but based on a GTO basis rather than a PW basis. The interface in CFOUR is based on a loose-coupling scheme facilitated by the light-weight MPI-based MiMiC communication library.

For AIMD and QM/MM-MD implementations using post-HF methods, the reduction of the computational cost is very important. We presented two methods that drastically reduce the computational cost without substantial loss of accuracy, namely a long-range coupling scheme and the use of a QM/QM MTS algorithm.

To verify the functionality and stability of our implementation, we performed simulations of gaseous water and a small and large liquid water system consisting of 1012 and 12001 water molecules, respectively, of which one molecule is treated by QM and the remainder by MM. The large water system was used for the investigation of the long-range electrostatic coupling. We showed that the energy fluctuations during an NVE simulation are very small and that the energy is conserved over time for all tested QM methods. The quality of the electrostatic coupling in QM/MM-MD simulations was investigated by comparison of RDFs, dipole moments, and vibrational frequencies. We showed that all of the mentioned properties agreed well with experimental data and with other theoretical studies.

With this coupling of CFOUR to MiMiC, we have an AIMD and QM/MM-MD implementation that is flexible in terms of the choice of QM method (e.g., DFT, MP2, CC, or CAS-SCF), and that has several methods to increase the computational efficiency of the simulations. The next step is to implement other quantum-chemical methods included in CFOUR like EOM-CC and to use the flexibility of the MiMiC framework to couple QM/MM-MD schemes with on-the-fly evaluation of molecular properties using a broad family of electronic-structure methods. Also, interfacing fast DFT-based AIMD codes to CFOUR, naturally opens to the further development of hierarchical QM/QM embedding schemes within MiMiC.

\section{Appendix}

\subsection{Multipole expansion for basis functions}

Here we give the explicit expressions for the multipoles used in eq. (\ref{multe}) and for the forces due to the long-range interactions. The integral form of the charge $C$, the dipole moment $\mu^\alpha$, the quadrupole moment $\Theta^{\alpha \beta}$, the octopole moment $\Omega^{\alpha \beta \gamma}$, and the hexadecapole moment $\Phi^{\alpha \beta \gamma \epsilon}$ of an electron density are

\be
C = \sum_{\mu, \nu} D_{\mu\nu}\langle \chi_\mu | \chi_\nu \rangle,
\ee

\be
\mu^\alpha = \sum_{\mu, \nu} D_{\mu\nu}\langle \chi_\mu | \alpha | \chi_\nu \rangle,
\ee

\be
\Theta^{\alpha \beta} = \sum_{\mu, \nu} D_{\mu\nu}\langle \chi_\mu | 3 \alpha \beta - |\br|^2\delta_{\alpha \beta}  | \chi_\nu \rangle,
\ee

\be
\begin{split}
\Omega^{\alpha \beta \gamma} & = \sum_{\mu, \nu} D_{\mu\nu}\langle \chi_\mu | 15 \alpha \beta \gamma - 3|\br|^2\left(\gamma \delta_{\alpha \beta} + \beta \delta_{\alpha \gamma} +\alpha \delta_{\beta \gamma} \right) | \chi_\nu \rangle,
\end{split}
\ee

and

\be
\begin{split}
\Phi^{\alpha \beta \gamma \epsilon}  & =
 \sum_{\mu, \nu} D_{\mu\nu}\langle \chi_\mu | 105 \alpha \beta \gamma \epsilon + 3 |\br|^4\left(\delta_{\alpha \beta} \delta_{\gamma \epsilon} + \delta_{\alpha \gamma} \delta_{\beta \epsilon} + \delta_{\alpha \epsilon} \delta_{\beta \gamma} \right) \\ 
& - 15 |\br|^2\left( \alpha \beta \delta_{\gamma \epsilon} + \alpha \gamma \delta_{\beta \epsilon} + \beta \gamma \delta_{\alpha \epsilon} + \alpha \epsilon \delta_{\beta \gamma} + \beta \epsilon \delta_{\alpha \gamma} + \gamma \epsilon \delta_{\alpha \beta} \right) | \chi_\nu \rangle.
\end{split}
\ee

The forces on the atoms are calculated via the negative derivative of the electrostatic potential $V_{iA}$ given in eq. (\ref{multe}) with respect to the coordinates of the long-range MM atoms $R_A$ and with respect to the coordinates of the QM atoms $R_I$ at which the basis functions are centered. The forces on the long-range MM atoms are given by

\be\label{lrmmf2}
\begin{split}
F_{iA}^\alpha(A) = -\frac{\delta V_{iA}}{\delta R_A^\alpha} & = q_A \Biggl[ C \frac{R_A^\alpha}{|\bR_A|^3} -   \mu^\alpha \frac{1}{|\bR_A|^3} + \sum_\beta \mu^\beta \frac{3R_A^\alpha R_A^\beta}{|\bR_A|^5} \\
 & - \sum_\beta \Theta^{\alpha \beta} \frac{R_A^\beta}{|\bR_A|^5} + \sum_{\beta, \gamma} \Theta^{\beta \gamma} \frac{2.5 R_A^\alpha R_A^\beta R_A^\gamma}{|\bR_A|^7} \\ 
 & - \sum_{\beta, \gamma} \frac{1}{2} \Omega^{\alpha \beta \gamma} \frac{R_A^\beta R_A^\gamma}{|\bR_A|^7} + \sum_{\beta, \gamma, \epsilon} \frac{7}{6} \Omega^{\beta \gamma \epsilon} \frac{R_A^\alpha R_A^\beta R_A^\gamma R_A^\epsilon}{|\bR_A|^9} \Biggr].
\end{split}
\ee 

The forces on the QM atoms at which the basis functions are centered are given by

\be\label{lrqmf2}
\begin{split}
F_{iA}^\alpha(I) = -\frac{\delta V_{iA}}{\delta R_I^\alpha} & = \sum_A q_A \Biggl[ C^{\bar{\alpha}} \frac{ 1}{|\bR_A|} + \sum_\beta \mu^{\beta,\bar{\alpha}} \frac{ R_A^\beta}{|\bR_A|^3} + \frac{1}{2} \sum_{\beta, \gamma} \Theta^{\beta \gamma, \bar{\alpha}} \frac{R_A^\beta R_A^\gamma}{|\bR_A|^5} \\
 & + \frac{1}{6} \sum_{\beta, \gamma, \epsilon} \Omega^{\beta \gamma \epsilon, \bar{\alpha}} \frac{R_A^\beta R_A^\gamma R_A^\epsilon }{|\bR_A|^7} + \frac{1}{24} \sum_{\beta, \gamma, \epsilon, \kappa} \Phi^{\beta \gamma \epsilon \kappa, \bar{\alpha}} \frac{ R_A^\beta R_A^\gamma R_A^\epsilon R_A^\kappa }{|\bR_A|^9}   \Biggr],
\end{split}
\ee

with $C^{\bar{\alpha}}$, $\mu^{\beta,\bar{\alpha}}$, $\Theta^{\beta \gamma, \bar{\alpha}}$, $\Omega^{\beta \gamma \epsilon, \bar{\alpha}}$, and $\Phi^{\beta \gamma \epsilon \kappa, \bar{\alpha}}$ the derivatives of charge, dipole moment, quadrupole moment, octopole moment, and hexadecapole moment with respect to the coordinates of the QM atoms $R_I^\alpha$. $C^{\bar{\alpha}}$ and $\mu^{\beta,\bar{\alpha}}$ are given by
\be
C^{\bar{\alpha}} = \frac{\delta C}{\delta R_I^\alpha} = \sum_{\mu, \nu} D_{\mu\nu} \left[ \left\langle \frac{\delta \chi_\mu}{\delta R_I^\alpha}  \bigg\vert \chi_\nu  \right\rangle + \left\langle \chi_\mu \bigg\vert \frac{\delta \chi_\nu}{\delta R_I^\alpha}  \right\rangle \right]
\ee
and
\be
\mu^{\beta,\bar{\alpha}} = \frac{\delta \mu^{\beta}}{\delta R_I^\alpha} = \sum_{\mu, \nu} D_{\mu\nu} \left[ \left\langle \frac{\delta \chi_\mu}{\delta R_I^\alpha}  \bigg\vert \beta \bigg\vert\chi_\nu  \right\rangle + \left\langle \chi_\mu \bigg\vert  \frac{\delta \beta}{\delta R_I^\alpha} \bigg\vert \chi_\nu  \right\rangle + \left\langle \chi_\mu \bigg\vert \beta \bigg\vert \frac{\delta \chi_\nu}{\delta R_I^\alpha}  \right\rangle \right].
\ee
The other multipole derivatives are calculated in an analogous manner.

\subsection{Energy conservation for MTS simulations}

\begin{figure}[h!]\centering
     \includegraphics[width=0.85\textwidth]{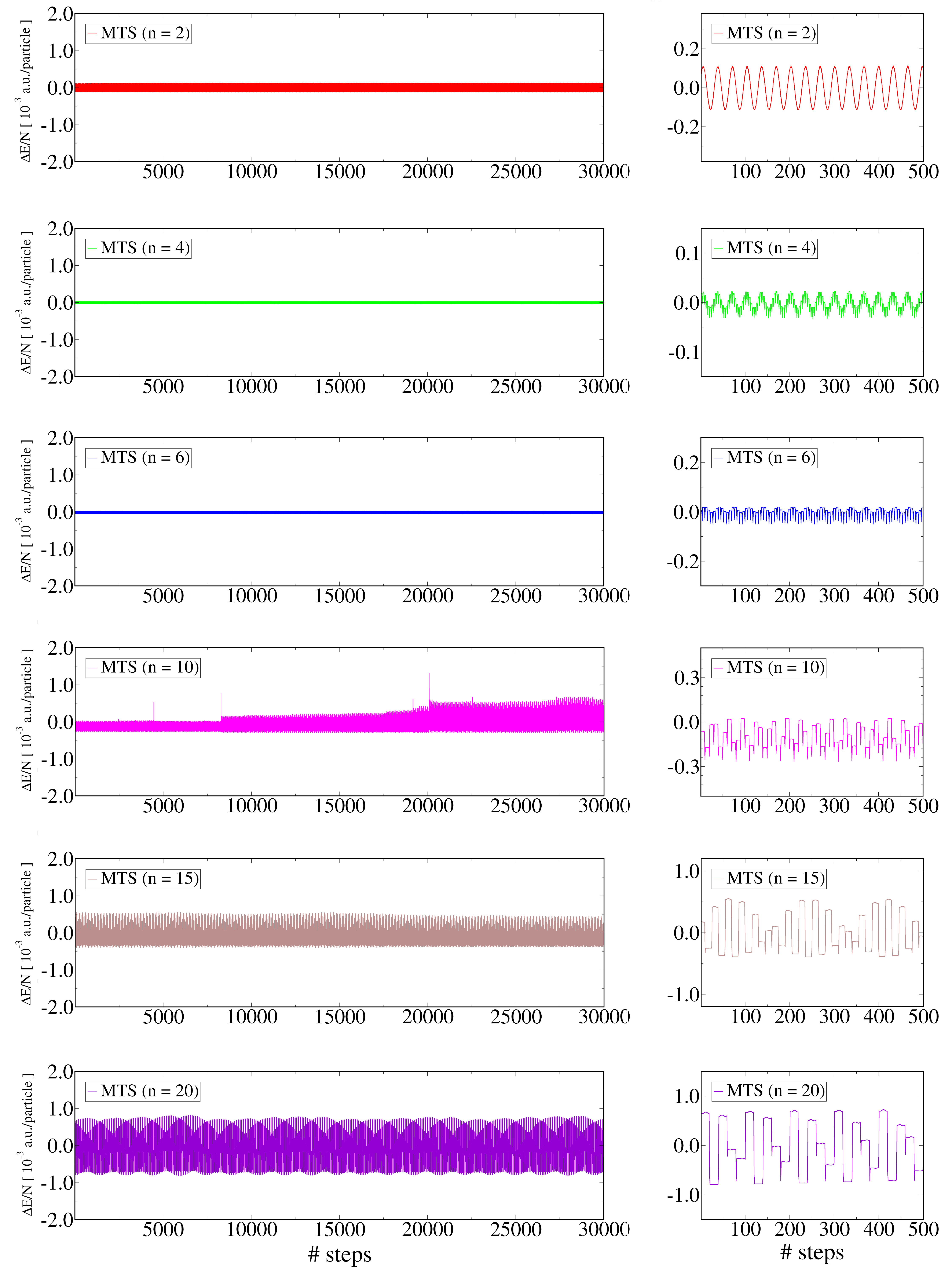}
    \caption{Fluctuation of the total energy per atom for hydrogen flouride in vacuum obtained from AIMD simulations at the combined CCSD(T)/cc-pVTZ and BLYP/PW levels of theory with CCSD(T)/cc-pVTZ calculations only every $n$th step where $n = 2, 4, 6, 10, 15,$ or $20$.}
    \label{nrg_mts}
\end{figure}

Figure (\ref{nrg_mts}) shows the energy conservation in AIMD simulations at the combined CCSD(T)/cc-pVTZ and BLYP/PW levels of theory for the different MTS factors. The fluctuations of the energy increase with increasing MTS factor. The stepped structure arise from the corrections steps which lead to jumps in the energy. However, there is no drift in energy seen during the simulations.

\section{Acknowledgement}
The authors acknowledge funding by the Deutsche Forschungsgemeinschaft
(DFG) within the project B5 of the TRR 146 (Project No.
233630050). Authors also acknowledge the support of the Research Council of Norway through the CoE Hylleraas Centre for Quantum
Molecular Sciences (Grant No. 262695), and the Norwegian
Supercomputing Program (NOTUR) (Grant No. NN4654K).
J.M.H.O. acknowledges financial support from VILLUM FONDEN (Grant No. 29478).

\bibliography{CFOUR_MIMIC_manuscript}

\end{document}